\documentstyle[multicol,pre,aps,psfig]{revtex}

\draft

\begin{document}

\title{Self-organized criticality and directed percolation}

\author{Alexei V\'azquez$^1$ and Oscar Sotolongo Costa$^{1,2}$}

\address{$^1$ Department of Theoretical Physics, Faculty of Physics, 
The University of Havana, Havana 10400, Cuba}

\address{$^2$ LCTDI, Faculty of Sciences, UNED, Madrid 28080, Spain}

\date{\today}

\maketitle

\begin{abstract}

A sandpile model with stochastic toppling rule is studied. The
control parameters and the phase diagram are determined through
a MF approach, the subcritical and critical regions are
analyzed. The model is found to have some similarities with
directed percolation, but the existence of different boundary
conditions and conservation law leads to a different
universality class, where the critical state is extended to a
line segment due to self-organization. These results are
supported with numerical simulations in one dimension. The
present model constitute a simple model which capture the
essential difference between ordinary nonequilibrium critical
phenomena, like DP, and self-organized criticality.

\end{abstract}


\begin{multicols}{2}

\section{introduction}

The idea of self-organized criticality (SOC) was introduced to
describe the behavior of a class of extended dissipative
dynamical systems which naturally evolve to a critical state,
consisting of avalanches propagating through the system
\cite{bak}. From the very beginning it was observed
that this new idea has some connections with ordinary critical
phenomena \cite{tang}. More recently a novel mean field (MF)
analysis of SOC was presented, which pointed out similarities
between SOC models and models with absorbing states
\cite{vespignani}. Directed percolation (DP) \cite{DP} is one of
the simplest and most recurrent models with absorbing states.
Under very general guidelines (locality, scalar variable, etc.)
it has been proposed that a wide range of models would fall into
DP universality class \cite{DP,janssen,grassberger}. Alhouh SOC
models do not belong to the DP universality class they have some
connection with DP \cite{olami,grassberger2,tadic}.

Recently, T\'adic and Dhar have shown that a class of stochastic
sandpile model has some analogy with DP \cite{tadic}. They
studied a directed sandpile model in which unstable sites
topples with probability $p$. They observed that above a
critical threshold $p_c$ the system shows SOC, while below the
system is not critical. The critical probability was identified
with the threshold for DP in a squared lattice and the scaling
exponents were obtained in terms of DP exponents. However, they
could not give a detailed description of the phase diagram of
the model, since their analysis was limited to the SOC regime
above $p_c$, while the state below $p_c$ could not be
characterized.

Following the work of T\'adic and Dhar we study a class of
stochastic sandpile models with undirected toppling rule. As in
their model, sites topples with a probability $p$ but now grains
are distributed to each nearest neighbor. In order to provide a
theoretical description of the model we have generalized the MF
theory by Vespignani and Zapperi \cite{vespignani} including the
control parameter $p$. In this way we obtain the complete phase
diagram of the model. The existence of a critical probability
$p_c$ and a quasi-stationary state below $p_c$ are obtained.
Based on the MF analysis and on the evolution rules we argue
that the state below $p_c$ is similar to DP, but with different
boundary conditions. Using this hypothesis we apply the scaling
theory developed for DP to the present stochastic sandpile
model. Numerical simulations in one dimension support our
hypothesis.

The paper is organized as follows. In section \ref{sec:MF} we
introduce the dynamical evolution rules for the stochastic
version of the Bak-Tang-Wiesenfeld (BTW) model.  We perform a
single site MF approximation and determine the average densities
in the stationary state. It is found that the driving rate $h$
and $p$ are the only control parameters and that the system is
critical in the line segment ($h=0^+$, $p_c\leq p\leq1$). Then
in section \ref{sec:scaling} we argument the connection with DP
and derive some scaling relations. In order to test our
predictions we have performed numerical simulations in one
dimension, the main results are presented in section
\ref{sec:simulations}. Finally, the summary and conclusions are
given in section \ref{sec:conclusions}.

\section{MF theory}
\label{sec:MF}

We study a stochastic sandpile model defined as follows. An
integer variable $z_i$ (height or energy) is assigned to each
site of a $d$-dimensional lattice and energy is added to the
system at rate $h$. When a site receives a grain and its energy
exceeds a threshold $z_c$ then, with probability $p$, it relaxes
according to the following rules $z_i\rightarrow z_i-g$ and
$z_j\rightarrow z_j+1$ at each of $g$ nearest neighbors.
Open boundary conditions are assumed. One may call this model
non-abelian sandpile model with stochastic rules. The non-abelian
behavior makes it different from other stochastic models such as
the Manna model \cite{manna}. However, we will simply call it
stochastic sandpile model.

The first step towards a comprehensive understanding of critical
phenomena is provided by mean-field (MF) theory, which gives
insight into the fundamental physical mechanism of the problem.
Thus, we start analyzing the stochastic sandpile model through a
MF approach. With this simple picture we introduce the
connection with directed percolation.

The first MF theory for sandpile models was introduced by Tang
and Bak \cite{tang}, and only deterministic toppling rules were
considered. Latter Caldarelli {\em et al} \cite{caldarelli}
generalize this MF theory to sandpile models with certain degree
of stochasticity in the toppling rules. In particular Caldarelli
{\em et al} studied sandpile models where $p$ is a function of
$z_i$, such that $0<p(z)<1$ below the critical threshold $z_c$
and $p(z)=1$ above. Their MF theory and further numerical
simulations reveals that the stochastic rules, introduced in
this way, changes the average of $z$ but does not destroy the
critical state \cite{caldarelli}. In the sandpile models
analyzed by Caldarelli {\em et al} $p$ is not a control
parameter, its average value is determined by the system
dynamics.  In these models the system self-organizes itself to a
stationary sate, where $\langle p(z)\rangle$ is such that the
system remains in a critical state. On the contrary in the
stochastic sandpile model considered here $p$ is a control
parameter. It is expected that for sufficiently small values of
$p$ the critical state will be destroyed. We have therefore to
develop a MF theory where $p$ appears explicitly as a control
parameter.

Recently Vespignani and Zapperi \cite{vespignani} have
introduced a more general framework. As a difference with
previous theories, their MF approach is not based in some
particular sandpile model but in general considerations which
are common to all of them, even extendible to other SOC models
\cite{vespignani}.  Within this formalism, SOC appears as a
special case of nonequilibrium critical phenomena. They have
divided the states each site can assume in stable ($s$),
critical ($c$), and active ($a$). Stable sites are those that
cannot become active by addition of energy, critical sites are
those that become active by addition of energy and active sites
are relaxing and transfer energy to their neighbors.  These
definitions becomes very clear in a deterministic sandpile
model.  For instance (assuming negligible the probability that a
site receives two energy grains) sites with $z<z_c-1$ are
stable, those with $z=z_c-1$ are critical and those with $z>z_c$
are active. However, in the stochastic model sites with $z\geq
z_c$ topple with probability $p$. Again sites with $z<z_c-1$ are
stable since they can never become active after receiving one
energy grain.  Nevertheless, those with $z\geq z_c-1$ have not a
well defined state. For instance, sites with $z=z_c-1$ may
topple after receiving one energy grain and, therefore, they
cannot be stable. However, they are not strictly critical
because only a fraction $p$ of them will topples after receiving
a grain of energy. Hence, in the case of stochastic sandpile
models the subdivision in stable, critical and active does not
covert all the possible states each site can assume, i.e.
$\rho_s+\rho_c+\rho_a<1$.

We divide the states each site can assume in stable ($s$),
unstable ($u$), and active ($a$) and denote their average
densities by $\rho_s$, $\rho_u$ and $\rho_a$, respectively. The
definition of stable and actives sites is the same considered by
Vespignani y Zapperi, while unstable sites are now those sites
that {\em may} become active by addition of energy.  Under these
definitions, sites with $z<z_c-1$ are stable, those with
$z=z_c-1$ are unstable and those with $z\geq z_c$ may be either
unstable or active. Only a fraction $p$ of the unstable sites
will become active after receiving energy and, therefore, are
critical sites, i.e.
\begin{equation} 
\rho_c=p\rho_u. 
\label{eq:0} 
\end{equation} 
This equality makes the connection between our MF approach and
that of Vespignani and Zapperi \cite{vespignani}. In the
deterministic limit $p=1$ there is no distintion between
critical and unstable sites.

\begin{figure}[t]\narrowtext
\centerline{\psfig{figure=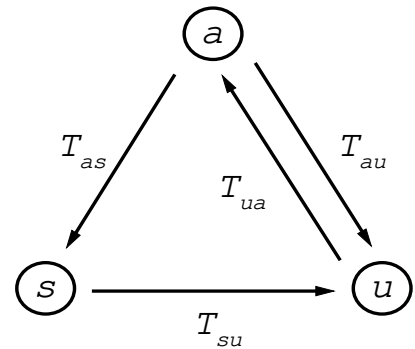,width=3.5in,height=3in}}
\caption{Schematic representation of the single site
approximation for the stochastic sandpile model. Sites are
divided in stable ($s$) unstable ($u$) and active ($s$).
$T_{mn}$ are the transition rates for the state $m$ to $n$.
Transitions which do not take place are not represented.}
\label{fig:0}
\end{figure}

To study the dynamics of the stochastic sandpile model we
consider the following Markov process for the average densities
\begin{equation} 
\frac{\partial}{\partial t}\rho_n = \sum_{m\neq n}T_{mn}\rho_m - 
\sum_{m\neq n}T_{nm}\rho_n,
\label{eq:1}
\end{equation} 
where $T_{nm}$ are the transition rates from the state $n$ to
the state $m$ (see fig. \ref{fig:0}). By definition of the
model, in one step stable sites never becomes active and
unstable sites never become stable, i.e.  $T_{sa}=T_{us}=0$.
$T_{as}=q$ and $T_{au}=1-q$, where $q$ is the fraction of active
sites that becomes stable after relaxing. In deterministic
models $q$ may be assumed equal to one \cite{vespignani}.
However in stochastic models $z$ may take values large enough
compared with $z_c$ in such a way that an active site main
become unstable after relaxing, i.e. $q<1$. Although an active
site with $z\gg z_c$ may remain active due to addition of energy
this type of transitions are of second order, they can be
neglected for $\rho_a$ and $h$ small.  The transition rates
$T_{su}$ and $T_{ua}$ depends on the probability per unit time
that a site receives energy. If $\rho_a$ and $h$ are small then
the probability per unit time that a site receives more than one
grain of energy is negligible, and the probability per unit time
that a site receive a grain of energy may be approximated by
\begin{equation} 
h_1=h+(g-\epsilon)\rho_a,  
\label{eq:2}
\end{equation} 
where $g-\epsilon$ is the effective number of nearest neighbors
and $\epsilon$ is the dissipation rate per toppling event, an
effective parameter which account for boundary dissipation.  If
$u$ ($p$) is the fraction of stable (unstable) sites that become
unstable (active) after receiving a grain of energy then
$T_{su}=uh_1$ ($T_{ua}=ph_1$).  Taking into account these
considerations the system of differential eqs. (\ref{eq:1}) is
reduced to
\begin{equation}
\frac{\partial}{\partial t}\rho_a =
-[1-(g-\epsilon)\rho_c]\rho_a +\rho_ch + O(h^2,h\rho_a),
\label{eq:3} 
\end{equation} 
\begin{equation}
\frac{\partial}{\partial t}\rho_s = q\rho_a -u(h+g\rho_a)\rho_s
+ O(h^2,h\rho_a),
\label{eq:4} 
\end{equation} 
together with the normalization condition
\begin{equation} 
\rho_s+\rho_u+\rho_a=1.  
\label{eq:5}
\end{equation} 
Notice that, among unstable sites, only the fraction of critical
sites $\rho_c=p\rho_u$ contributes to the system dynamics, the
other fraction is only relevant through the normalization
condition in eq. (\ref{eq:5}). The system of equations is
completed by the equation of energy balance
\begin{equation}
\frac{\partial}{\partial t}E = (h - \epsilon\rho_a) L^d, 
\label{eq:5a}
\end{equation} 
where $E$ is the total energy of the system, $hL^d$ is the
average influx of energy and $\epsilon\rho_a L^d$ the average
outflux of energy.

\subsection{Critical state}

In the stationary state ($\frac{\partial\rho_n}{\partial t}=0$,
$\frac{\partial E}{\partial t}=0$) from eqs.
(\ref{eq:3}-\ref{eq:5a}) and (\ref{eq:0}) we obtain
\begin{equation}
\rho_a = \frac{h}{\epsilon},\ \ \ \ \rho_c=\frac{1}{g} + O(h),
\label{eq:6}
\end{equation}
\begin{equation}
\rho_u=\frac{1}{pg} + O(h),\ \ \ \ \rho_s=\frac{pg-1}{pg} + O(h),
\label{eq:7}
\end{equation}
\begin{equation}
\frac{q}{u}=\frac{pg-1}{p} + O(h).
\label{eq:8}
\end{equation}
Comparing this expressions with the ones obtained by Vespignani
and Zapperi we observe that the average densities of active and
critical sites have the same stationary solutions. The
differences appear in the density of stable and unstable sites,
which now depends on the new control parameter $p$.

For $1/g\leq p\leq1$ we have $1/g\leq\rho_u\leq1$,
$0\leq\rho_s\leq(g-1)/g$ and, therefore, there is not any
inconsistency in the stationary solutions obtained above. In
this range of $p$, within the MF approach, there is no
distinction between the critical state of stochastic and
deterministic sandpile models. The model is critical in the
double limit $h,\epsilon\rightarrow0$ and
$h/\epsilon\rightarrow0$ \cite{vespignani} and the
susceptibility,
\begin{equation}
\chi=\frac{\partial\rho_a}{\partial h} = \frac{1}{\epsilon},
\label{eq:9}
\end{equation}
diverges in the critical state. For a small perturbation around
the subcritical state $\rho_a=h/\epsilon+\Delta\rho_a$,
$\rho_c\approx1/g$ and from eq.  (\ref{eq:3}) one obtains
\begin{equation}
\Delta\rho_a(t)\propto\exp(-\epsilon t/g),
\label{eq:10}
\end{equation}
in agreement with the result obtained for deterministic models
\cite{vespignani}.

From eqs. (\ref{eq:9}) and (\ref{eq:10} one may think that
$\epsilon$ is a control parameter of the model. However, if the
dissipation takes place only at the boundary then $\epsilon$
will decrease with decreasing system size, because the number of
actives sites in the bulk grows faster than the number of active
sites at the boundary. Hence, we are just dealing with a finite
size effect. The statement the system is not in a critical state
is equivalent to the statement there is no critical state in a
finite system. If dissipation takes place only at the boundary
$\epsilon$ is not a control parameter, it just reflects a finite
size effect which at the same time is a necessary condition to
obtain a stationary state. In this sense the criticality here is
different from the criticality at phase transitions where
boundary effects always disappear in the thermodynamic limit
\cite{bak}.

On the contrary, the driving field $h$ is actually a control
parameter.  Since $h$ must satisfy $h<\epsilon$ and
$\epsilon\rightarrow 0$ when $L\rightarrow\infty$ then we must
fine tune $h$ to zero in order to obtain criticality in the
thermodynamic limit. The time scale separation becomes a
necessary condition for criticality.  Now if we assumes
separation of time scales then $p$ will be the only control
parameter of the model. This hypothesis is in general fulfilled
in computer simulations, where a new grain of energy is added
only once there is no active site.

\subsection{Break-down of SOC by stochastic rules}

When $0<p<1/g$ the stationary solutions in eqs.
(\ref{eq:6}-\ref{eq:8}) are no longer valid, because they imply
$\rho_u>1$ and $\rho_s<0$. To understand the origin of this
inconsistency let us analyze the variation of $\rho_u$ and
$\rho_c$ with $p$. In the deterministic case $p=1$ there is no
distinction between unstable and critical sites, i.e.
$\rho_c=\rho_u=1/g$.  However, when $p<1$ critical sites are a
fraction $\rho_c=p\rho_u$ of unstable sites. Hence, since
$\rho_c=1/g$ in the stationary state, the system has to
self-organize itself increasing the average density of unstable
sites to $\rho_u=1/pg>1/g$ and decreasing the average density of
stable sites. But when $p=1/g$ we have $\rho_u=1$ and $\rho_s=0$
and, therefore, if we continue decreasing $p$ the system cannot
provide more unstable sites. Then $\rho_c<1/g$ and the
stationary solutions in eqs. (\ref{eq:6}-\ref{eq:8})
break-downs.

Let us assume that for $0<p<1/g$ the average densities reache an
stationary state, which off course cannot be given by eqs.
(\ref{eq:6}-\ref{eq:8}).  From eq.  (\ref{eq:3}) it results that
\begin{equation}
\rho_a = \frac{\rho_c h}{1-(g-\epsilon)\rho_c},
\label{eq:11}
\end{equation}
Substituting this expression in eq. (\ref{eq:5a}) one obtains
\begin{equation}
\frac{\partial}{\partial t}E=\frac{1-g\rho_c}{1-(g-\epsilon)\rho_c}hL^d.
\label{eq:12}
\end{equation}
According to eq. (\ref{eq:0}) $\rho_c=p\rho_u\leq p<1/g$,
independent of $\rho_u$.  Hence, $\frac{\partial E}{\partial
t}>0$ and the total energy will increase linearly with time.
Moreover, stable sites are those with $z<z_c-1$ and, therefore,
it is expected that after a time long enough there will be no
stable sites. In this quasi-stationary state the energy
increases with time and the average densities will take the
stationary values
\begin{equation}
\rho_a = \frac{hp}{1-(g-\epsilon)p},\ \ \ \ \rho_c=p + O(h),
\label{eq:13}
\end{equation}
\begin{equation}
\rho_u=1 + O(h),\ \ \ \ \rho_s=0
\label{eq:14}
\end{equation}
\begin{equation}
q=0.
\label{eq:15}
\end{equation}
Now it is clear that $p$ is a control parameter of the class of
stochastic sandpile models analyzed here.

The susceptibility in this region is given by, assuming
$\epsilon\ll g$,
\begin{equation}
\chi = \frac{p}{1-pg}.
\label{eq:16}
\end{equation}
For a small perturbation around the subcritical state we have
$\rho_a=\chi h+\Delta\rho_a$ and $\rho_c\approx p$, then from eq.
(\ref{eq:3}) one obtains, again considering $\epsilon\ll g$,
\begin{equation}
\Delta\rho_a(t)\propto\exp[-(1-pg)t].
\label{eq:17}
\end{equation}
The critical state breaks-down by the stochastic rules, once
$p<p_c$. To reach the critical state we have to fine tune $p$.
Near the critical threshold $1/g$ the susceptibility
$\chi\sim(p_c-p)^{-1}$ and the characteristic time
$\sim(p_c-p)^{-1}$ diverges.

In summary, we found a critical probability $p_c$ above which
the system is in a SOC state, while it is in a subcritical state
below.

\begin{figure}[t]\narrowtext
\centerline{\psfig{figure=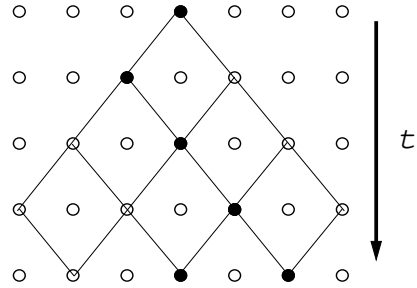,width=3.5in,height=3in}}
\caption{Evolution of an avalanche in the stochastic sandpile
model. Empty sites are inactive (stable+unstable) and filled
sites are active. This figure clearly shows that the avalanche
evolves in a directed square lattice, where the probability
that a site is present is $\rho_c$.}
\label{fig:1}
\end{figure}

\section{Scaling theory}
\label{sec:scaling}

$\rho_c$ is the probability that a site becomes active after
receiving a grain of energy. In the absence of an external field
it is the probability that a site becomes active if one of its
neighbors was active in the previous step. This problem is
equivalent to site directed percolation in $d+1$ dimensions ($d$
spatial dimensios$+$time), $\rho_c$ being the probability that a
site is present. The only different is found in the boundary
conditions, while in DP the system is assumed infinite here we
deal with a finite system with open boundaries. This picture is
better represented in fig. \ref{fig:1}.

\subsection{$0<p<p_c$}

According to the MF theory below the $p_c$ there are no stable sites
and the fraction of critical sites is given by $\rho_c=p$. The
evolution rules are thus that of DP. The existence of different
boundary conditions may carry as a consequence that some scaling
exponents result different, however the nature of the phenomena
is the same. For instance, DP near a wall reveals that the
correlation length exponents are identical to those obtained in
DP in an infinite lattice \cite{essam0}, but other exponents
take different values. This is a consequence of the fact that in
DP near a wall the avalanches are a subset of the avalanches in
DP in an infinite lattice. We thus expect a similar behavior in
the stochastic sandpile model below $p_c$. In this case,
avalanches starting far from the boundary behaves as in DP in an
infinite lattice, while avalanches starting near the boundary
behaves like the avalanches in DP near a wall. Hence, the
correlation lengths and the correlation length exponents are
identical to those of DP in an infinite lattice, i.e.
\begin{equation}
\xi_\bot\sim(p_c-p)^{-\nu_\bot},\ \ \ \ \xi_{||}\sim(p_c-p)^{-\nu_{||}}.
\label{eq:30}
\end{equation}
where $\xi_\bot$ and $\xi_{||}$ are the spatial and temporal
correlation lengths, respectively, and $\nu_\bot$ and $\nu_{||}$
the correlation length exponents.

On the other hand, based on this analogy with DP we write the
following scaling relation for the average density of active
sites at site $x$ and time $t$, given a site was active in the
origin $x=0$ at $t=0$,
\begin{equation}
\rho_a(x,t)=t^{\eta-\frac{2}{z}}f\biggl(\frac{x^2}{t^{2/z}},
\frac{t}{\xi_{||}}\biggr),
\label{eq:31}
\end{equation}
first introduced by Grassberger and de la Torre in the contest
of DP \cite{grassberger3}. Here $\eta$ is a scaling exponent and
$z$ the dynamic scaling exponent, as it is usually defined in
the contest of critical phenomena. Moreover, the probability
that the avalanche survive up to time $t$ is given by
\begin{equation}
P(t)=t^{-\delta}g\biggl(\frac{t}{\xi_{||}}\biggr),
\label{eq:32}
\end{equation}
where $\delta$ is another scaling exponent. From eq.
(\ref{eq:31}) one can derive other scaling laws for the average
number of active sites $n(t)$, the cluster mass $m(t)$ and the
mean squared displacement $R^2(t)$ at time $t$, resulting
\begin{equation}
\begin{array}{l}
n(t)=\int
d^dx\rho_a(x,t)=t^{\eta}f_1\bigl(\frac{t}{\xi_{||}}\bigr),\\
m(t)=\int_0^tdt^\prime
n(t^\prime)=t^{1+\eta}f_2\bigl(\frac{t}{\xi_{||}}\bigr),\\
R^2(t)=\frac{1}{n(t)}\int
d^dx\rho_a(x,t)x^2=t^{2/z}f_3\bigl(\frac{t}{\xi_{||}}\bigr).
\end{array}
\label{eq:33}
\end{equation}

The exponent $z$ is not independent, since $t\sim r^z$ from eq.
(\ref{eq:30}) one obtains
\begin{equation}
z=\frac{\nu_{||}}{\nu_\bot},
\label{eq:34}
\end{equation}
and it is therefore identical to that of DP in an infinite
lattice. Nevertheless, the exponents $\eta$ and $\delta$ depend
on the boundary conditions, as it is observed in DP near a wall
\cite{essam0}.

\subsection{$p_c\leq p<1$}

We assume that the scaling laws in eqs.
(\ref{eq:31}-\ref{eq:33}) are also valid above $p_c$, but with
$\xi_{||}=\xi_{||}(L)$. In this region the dynamical evolution
is independent of $p$ and the characteristic length and time
depends only on the lattice size $L$, according to
\begin{equation}
\xi_\bot\sim L,\ \ \ \ \xi_{||}\sim L^z.
\label{eq:35}
\end{equation}
However, as it is shown below, for $p>p_c$ the global
conservation introduces a constraint between the exponents
$\eta$ and $z$.

Let us calculate the average flux of energy $J(r)$ outside an
sphere of radius $r$, given a grain of energy was added at the
origin $r=0$ at $t=0$. The energy flux is proportional to
the gradient of the average density of active sites and, therefore,
\begin{equation}
J(r)\propto r^2\int dt\frac{\partial}{\partial r}\rho_a(r,t).
\label{eq:36}
\end{equation}
Substituting the scaling relation for $\rho_a(r,t)$
(\ref{eq:31}) in this expression it results that
\begin{equation}
J(r)=r^{(1+\eta)z-2}f_4\biggl(\frac{r}{\xi_\bot}\biggr).
\label{eq:37}
\end{equation}
Now, conservation implies that $J(r)=1$ for $r<\xi_\bot\sim L$
and, therefore,
\begin{equation}
(1+\eta)z=2.
\label{eq:38}
\end{equation}
This scaling relation may seem unusual, a more familiar
expression is obtained if one calculate the mean avalanche size
\begin{equation}
\langle s\rangle =\int dt n(t)\sim L^{(1+\eta)z}\sim L^2.
\label{eq:39}
\end{equation}
This scaling relation was previously obtained by Dhar
\cite{dhar} but for a particular sandpile model. We have here
demonstrated, using scaling arguments, that it holds for any
sandpile model with global conservation.

\subsection{Directed models}

In the directed stochastic models there is a preferent direction
$l$ for the avalanche evolution. This preferent spatial
direction may be identified with the time direction in
undirected models and then apply the scaling theory developed
above. This means that the scaling relations in eqs.
(\ref{eq:31}-\ref{eq:33}) are also valid for directed
models, but now $x$ is a ($d-1$)-dimensional vector in the space
of the non-preferent directions and $t$ gives the evolution in
the preferent direction $l$.

Another important difference between undirected and directed
models is the place where topling take places during the
evolution of the avalanche. In the undirected model not only the
sites in the avalanche front but also sites inside this front may
be active, transfering energy to their neighbors. On the
contrary in directed models all active sites are in the
avalanche front, i.e. at step $t$ all active sites are in the
layer $l=t$.  Moreover, actives sites in layer $l$ transfer
energy only to those neighbors in layer $l+1$. Hence, in
directed models the energy flux take places following the
preferent direction, while it take places in all directions in
the undirected case.

An inmediate consequence of this difference is that in directed
models the average outflux of energy  from the $l=t$ to the
$l+1$ layer, which is proportional to the average number of
active sites in the $l=t$ layer, equals one and, therefore,
\begin{equation}
n(t)\sim 1.
\label{eq:40c}
\end{equation}
Then, form eq. (\ref{eq:33}) one obtains that
\begin{equation}
\eta=0.
\label{eq:40d}
\end{equation}
The energy balance thus leads to a different constraint for the
set of scaling exponents $(\eta,z,\delta)$, i.e. the directed
model is in a diferent universality class.

\section{Numerical simulations and discussion}
\label{sec:simulations}

In order to test our predictions we have performed numerical
simulations of the stochastic sandpile model in one dimension.
We have investigated both regimes of the phase diagram, the
region similar to DP below $p_c$ and the SOC region above. In
both regions we start with a flat pile, i.e. zero height in all
sites, and let the system evolve to the stationary state. Above
$p_c$ the stationary state is characterized by a constant
average energy per site, which was taken as the stationary
condition. Below $p_c$ the energy increases linearly with time
and, therefore, we look for another criterion of stationarity.
According to the MF theory in the quasi-stationary state below
$p_c$ we have $\rho_u=1$, which was taken as the stationary
condition. In all cases we start measuring after the system
reached the stationary state. Average where taken over 10 000
000 avalanches below $p_c$ and over 1000 000 avalanches above.
Below $p_c$ we use the lattice size $L=10240$, which was large
enough to avoid finite size effects for the values of $p$
considered. Above $p_c$ we use $p=0.708$ and different lattice
sizes.

To obtain an estimate of $p_c$ we have calculated the
correlation lengths in the subcritical state for different
values of $p$ and fitted the numerical data to the scaling laws
in eq. (\ref{eq:30}). These magnitudes where computed in the
simulations using the following expressions
\begin{equation}
\begin{array}{l}
\xi_\bot^2\sim \frac{\sum_{t=0}^\infty\sum_{i=0}^L(i-i_0)^2\rho_{ai}}
{\sum_{t=0}^\infty\sum_{i=0}^L\rho_{ai}},\\
\\
\xi_{||}\sim \frac{\sum_{t=0}^\infty\sum_{i=0}^Lt\rho_{ai}}
{\sum_{t=0}^\infty\sum_{i=0}^L\rho_{ai}}
\end{array}
\label{eq:41}
\end{equation}
where $i_0$ is the position of the initial active site, $t$ is
the number of steps measured in the time scale of the avalanche
and $\rho_{ai}=1$ ($\rho_{ai}=0$) in active (unstable) sites.

\begin{figure}\narrowtext
\centerline{\psfig{figure=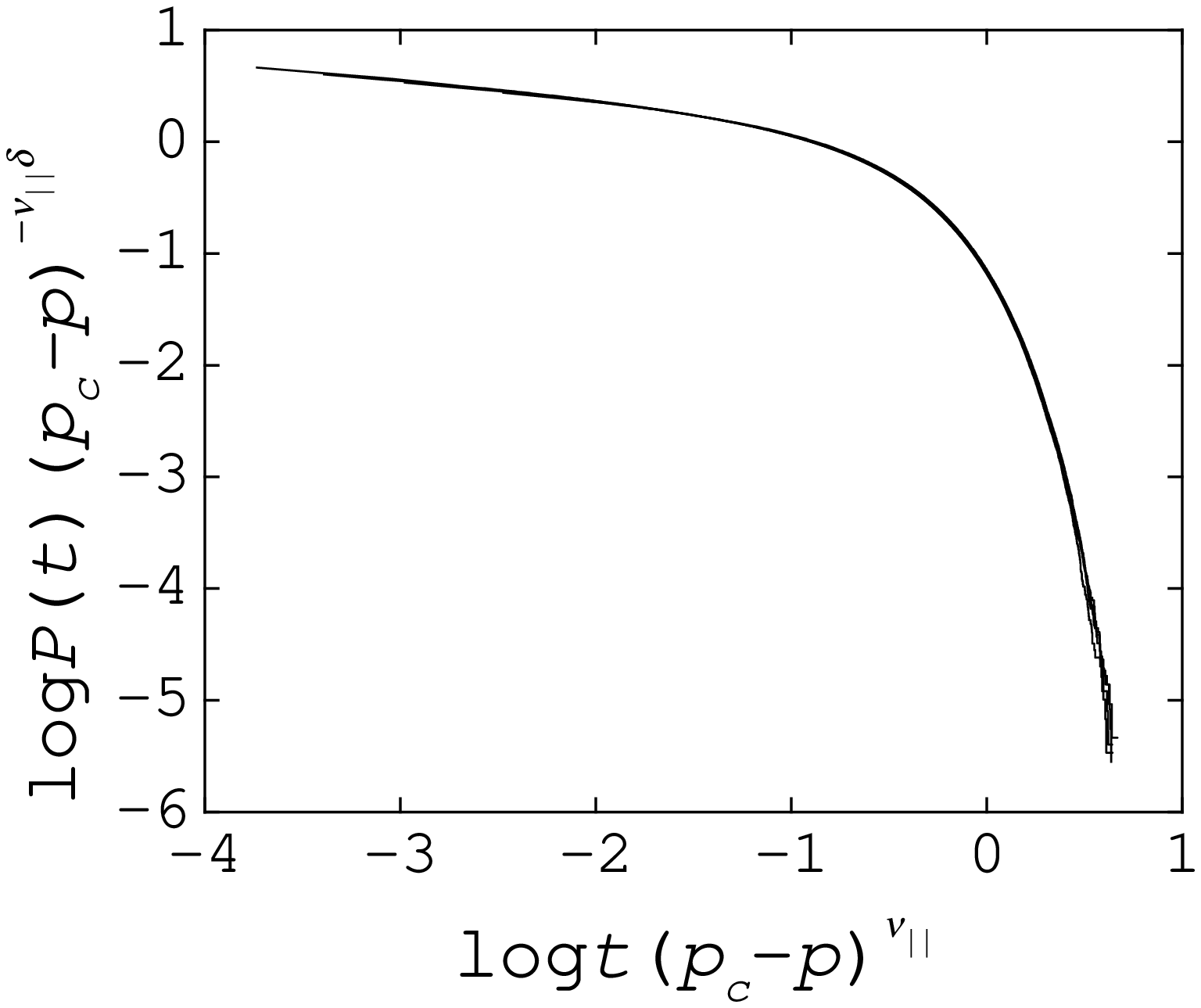,width=3.5in,height=3in}}
\caption{Data collapse plot for $P(t)$ in the subcritical
state for $p=0.670$, 0.688, 0.696 and 0.700.}
\label{fig:3}
\end{figure}

The log-log plot of the correlation lengths versus $p_c-p$ is
shown in fig. \ref{fig:3}. The best fit to the numerical data
was obtained for
\begin{equation}
\begin{array}{ll}
p_c=0.707\pm0.002, & (0.705485);\\
\nu_\bot=1.07\pm0.03, & (1.0968);\\
\nu_{||}=1.71\pm0,03, & (1.7338).
\end{array}
\label{eq:42}
\end{equation}
Enclosed in parenthesis are the series expansion estimates for
DP in an infinite lattice reported in \cite{jenssen}. Within the
numerical error there is a complete agreement between the values
reported here and those of DP.

Then we proceed to determine the exponents $\delta$, $\eta$ and
$z$ from the data collapse plots of $P(t)$, $m(t)$ and $R^2(t)$,
using the scaling laws in eq. (\ref{eq:33}). The corresponding
plots are sown in figs. \ref{eq:3}-\ref{eq:5}. The best data
collapse was obtained for
\begin{equation}
\begin{array}{ll}
\delta=0.18\pm0.01, & (0.15947);\\
\eta=0.27\pm0.01, & (0.31368);\\
z=1.59\pm0.01, & (1.58074);\\
\nu_{||}=1.73\pm0.01, & (1.7338);\\
(1+\eta)z=2.02\pm0.02.
\end{array}
\label{eq:44}
\end{equation}
From the data collapse we have obtained a better estimate for
$\nu_{||}$ and, using the scaling relation (\ref{eq:34}) and the
value of $z$ in (\ref{eq:44}), we obtain the better estimate for
$\nu_\bot$
\begin{equation}
\begin{array}{ll}
\nu_\bot=1.09\pm0.02, & (1.0968).
\end{array}
\label{eq:45}
\end{equation}
As it was expected the critical probability, the correlation
length exponents, and $z$ are identical, within the numerical
error, to those reported for DP in an infinite lattice, while
$\eta$ and $\delta$ results different.

\begin{figure}\narrowtext
\centerline{\psfig{figure=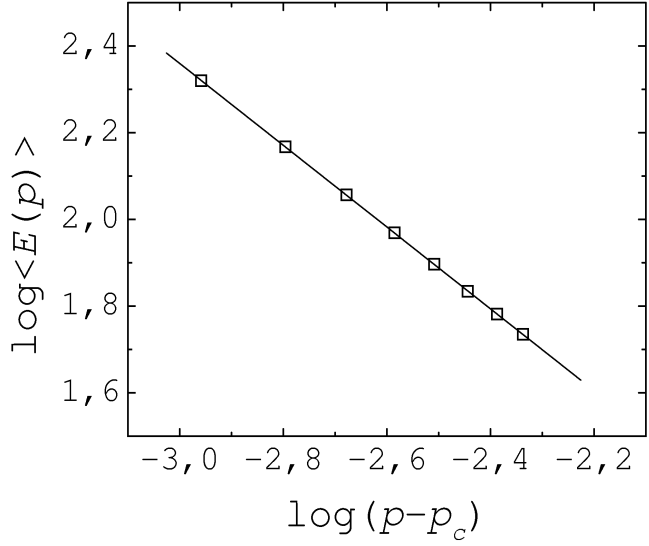,width=3.5in,height=3in}}
\caption{Average energy per lattice site as a function of $p$ in
the SOC state. The continuous line is a linear fit in the
log-log scale.}
\label{fig:3a}
\end{figure}

Now let us analyze the numerical simulations in the SOC region
$p_c\leq p<1$. In this case we can obtain an estimate of the
critical probability from the divergence of the average energy
per site $\langle E\rangle$ near $p_c$. In the SOC region
$\langle E\rangle$ reaches a stationary value but it increases
with time below $p_c$. One thus expect that $\langle E\rangle$
diverges when the system approaches the critical probability
from above. We observe that the divergence of $\langle E\rangle$
can be fitted to the power law dependency $\langle
E\rangle\sim(p-p_c)^{-\lambda}$, where $\lambda$ is a scaling
exponent. In fig. \ref{fig:3a} we have plot the best fit to the
numerical data for a lattice size $L=1280$, which was obtained
for
\begin{equation}
\begin{array}{ll}
p_c=0.704\pm0.01, & (0.705485),
\end{array}
\label{eq:46}
\end{equation}
which is close to the DP value.

\begin{figure}\narrowtext
\centerline{\psfig{figure=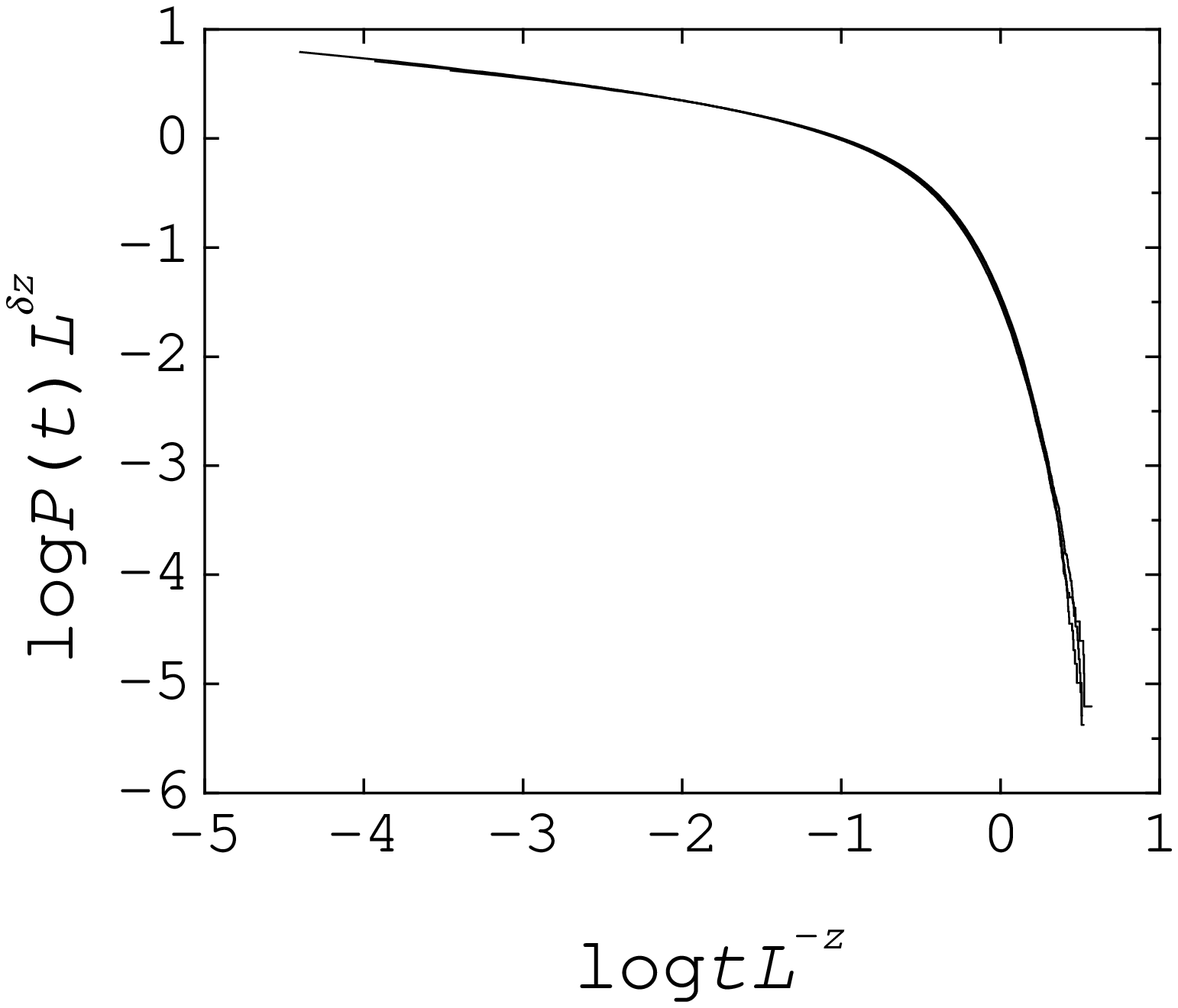,width=3.5in,height=3in}}
\caption{Data collapse plot for $P(t)$ in the SOC state for
$L=160$, 320 and 640.}
\label{fig:6}
\end{figure}

\begin{figure}\narrowtext
\centerline{\psfig{figure=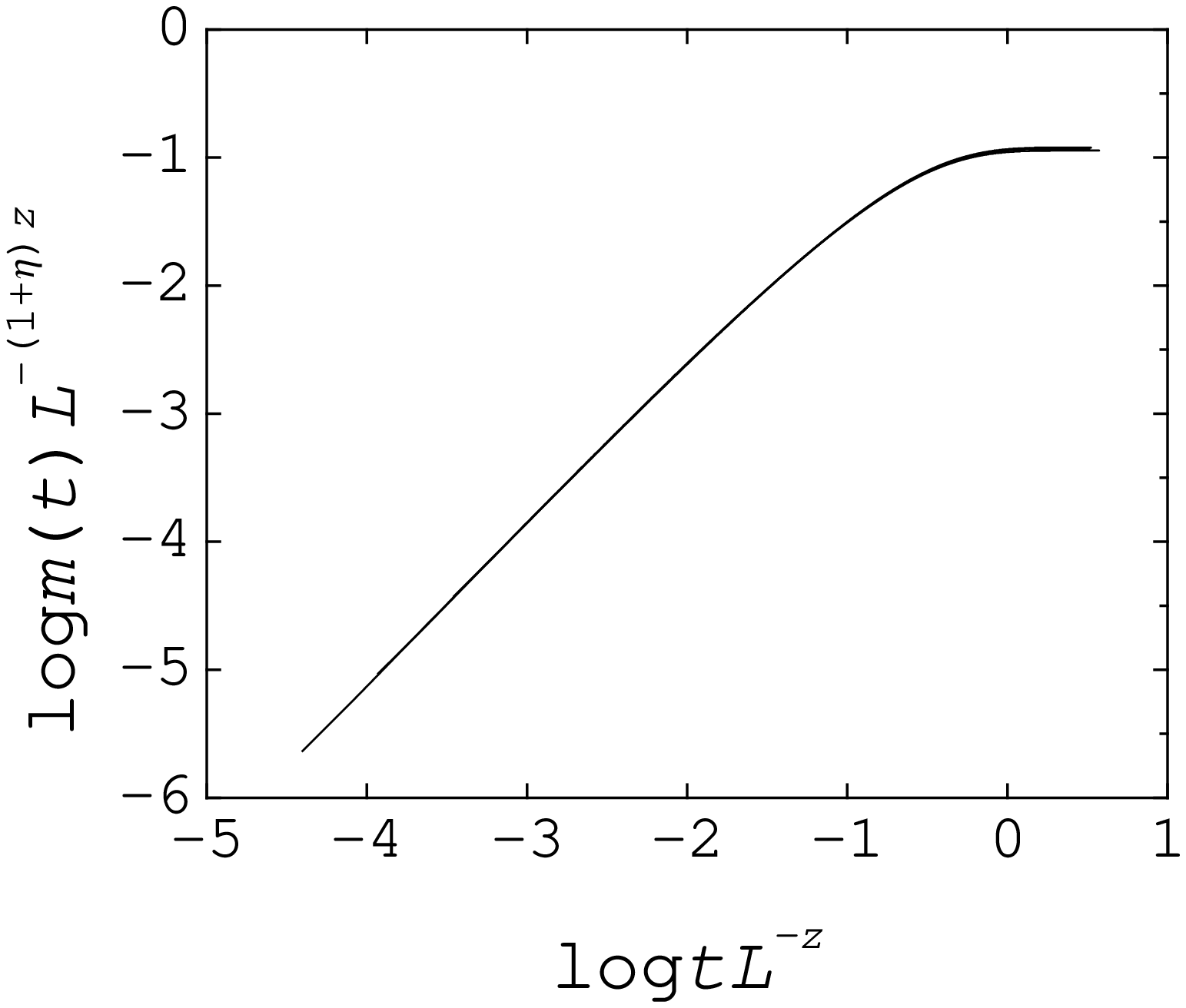,width=3.5in,height=3in}}
\caption{Data collapse plot for $m(t)$ in the SOC state for
$L=160$, 320 and 640.\\}
\label{fig:7}
\end{figure}

\begin{figure}\narrowtext
\centerline{\psfig{figure=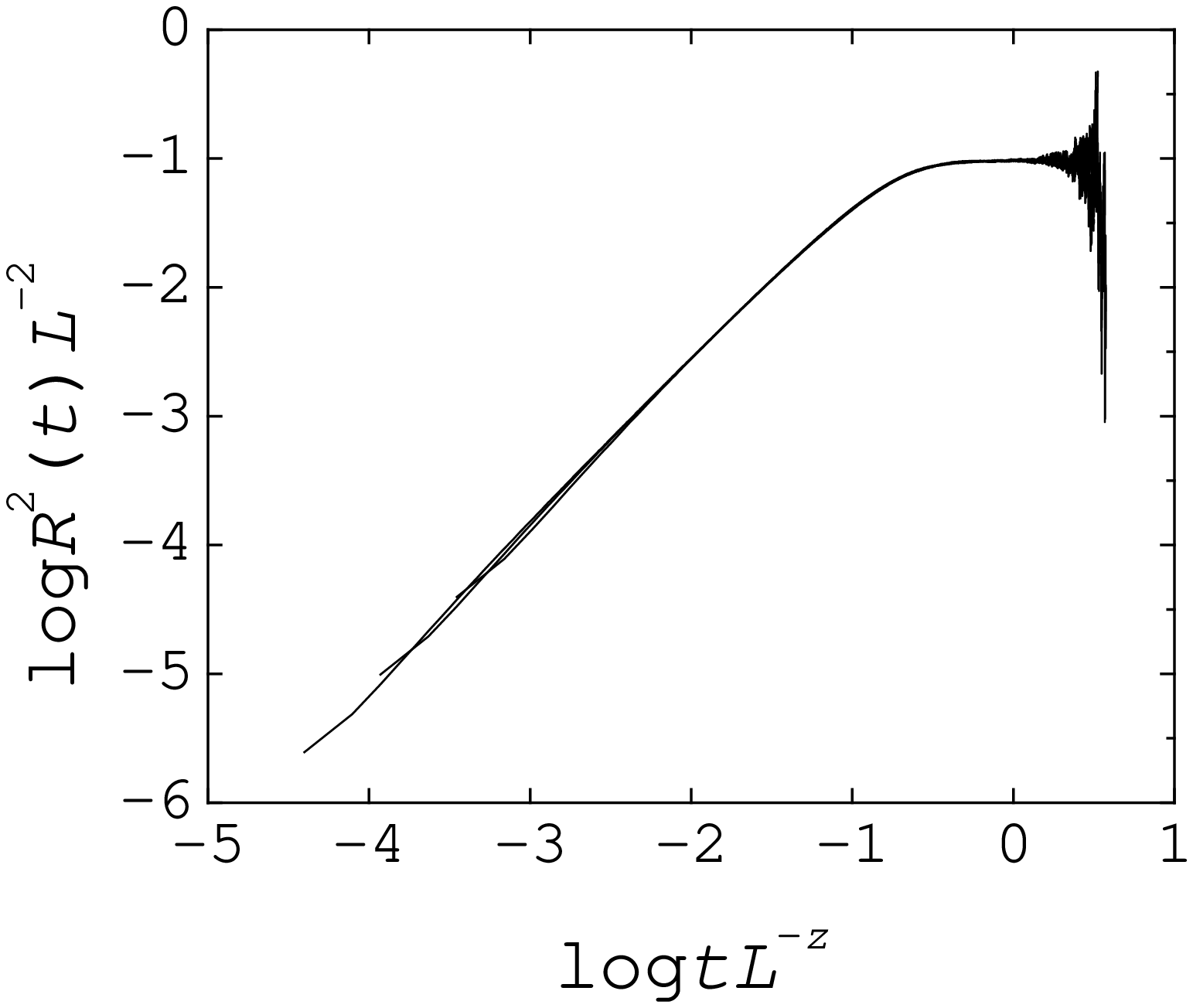,width=3.5in,height=3in}}
\caption{Data collapse plot for $R^2(t)$ in the SOC state for
$L=160$, 320 and 640.}
\label{fig:8}
\end{figure}

Then we proceed to determine the exponents $\delta$, $\eta$ and
$z$ from the data collapse above $p_c$, using the scaling laws
in eqs. (\ref{eq:33}) and (\ref{eq:35}). The best data collapse
is shown in figs. \ref{fig:6}-\ref{fig:8} with
\begin{equation}
\begin{array}{ll}
\delta=0.18\pm0.01, & (0.15947);\\
\eta=0.28\pm0.01, & (0.31368);\\
z=1.57\pm0.01, & (1.58074);\\
(1+\eta)z=2.01\pm0.02
\end{array}
\label{eq:47}
\end{equation}
From the comparison of these values with those in eq.
(\ref{eq:44}) we conclude that the scaling exponents $\delta$,
$\eta$ and $z$ are the same above and below $p_c$. Moreover, the
scaling relation in eq. (\ref{eq:38}) is in both cases
satisfied, although it was demonstrating only for $p>p_c$.
Hence, there are only three independent scaling exponents,
$\nu_\bot$, $\nu_{||}$ and $\delta$, while $z$ and $\eta$ can be
determined using the scaling relations in eqs. (\ref{eq:34}) and
(\ref{eq:38}). The correlation length exponents are identical to
those of DP in an infinite lattice while $\delta$ depends on the
boundary conditions and, therefore, changes the universality
class.

\section{Summary and conclusions}
\label{sec:conclusions}

We have obtained, through a MF analysis, the phase diagram of
the stochastic sandpile model. There is a critical probability
$p_c$ above which the system is in a SOC state, where the
correlation lengths diverge in the thermodynamic limit. Below
$p_c$ the system is subcritical, it is characterized by a finite
susceptibility which diverges when the critical state is
approached. While the stationary state in the SOC state is
characterized by a well defined average energy per lattice site
the subcritical state is not completely stationary, since the
average energy per lattice size increases linearly with time. It
was then corroborated that the global conservation is a
necessary condition to obtain SOC in sandpile models.

Using scaling arguments it was demonstrated that in the
subcritical region the stochastic sandpile model is similar to
DP, but with different boundary conditions. On the other hand,
the scaling theory in the SOC state reveals that global
conservation introduces a constraint among the scaling
exponents, generalizing previous results obtained for particular
sandpile models. We have provided a general demonstration of the
scaling law $\langle s\rangle\sim L^2$.

Numerical simulations have corroborated the predictions of the
MF and scaling theory. The correlation length exponents and the
critical probability were found, within the numerical error,
identical to the estimates for DP. However, the existence of
different boundary conditions and conservation law carries as a
consequence that other exponents result different, changing the
universality class.

We must emphasize that the stochastic sandpile model is not just
another cellular automaton showing SOC, but a very nice example
to understand the differences and similarities between SOC and
ordinary nonequilibrium critical phenomena. The comparison of
the phase diagram of this model with that of DP reveals the
essential property of SOC, the insensitive to changes in certain
''control'' parameter, which is off course no more a control
parameter. While the critical state in DP is restricted to a
point in the phase diagram, in the stochastic sandpile model it
is extended through a line segment.

\section*{Acknowledgments}

This work was partially supported by the {\em Alma Mater} prize,
given by The University of Havana. We thanks Alessandro
Vespignani for helpful comments and discussion during the
preparation of this manuscript. The numerical simulations where
performed using the computer resources of The Abdus Salam ICTP,
during the visit of A. V\'azquez to this center under the 1998
Federation Arrangement.

\newpage

\begin{figure}\narrowtext
\centerline{\psfig{figure=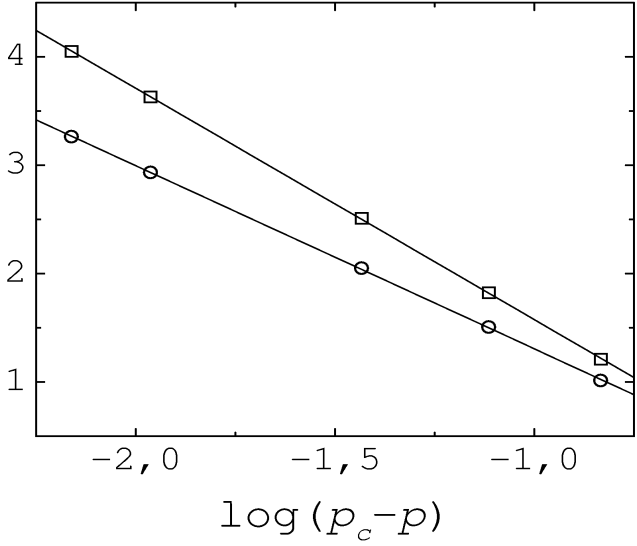,width=3.5in,height=3in}}
\caption{The correlation lengths $\log\xi_\bot$ (squares) and
$\log\xi_{||}$ (circles) as a function of $p$ in the subcritical
state. The lines are linear fits to the log-log plot.}
\label{fig:2}
\end{figure}

\vskip 0.5in

\begin{figure}\narrowtext
\centerline{\psfig{figure=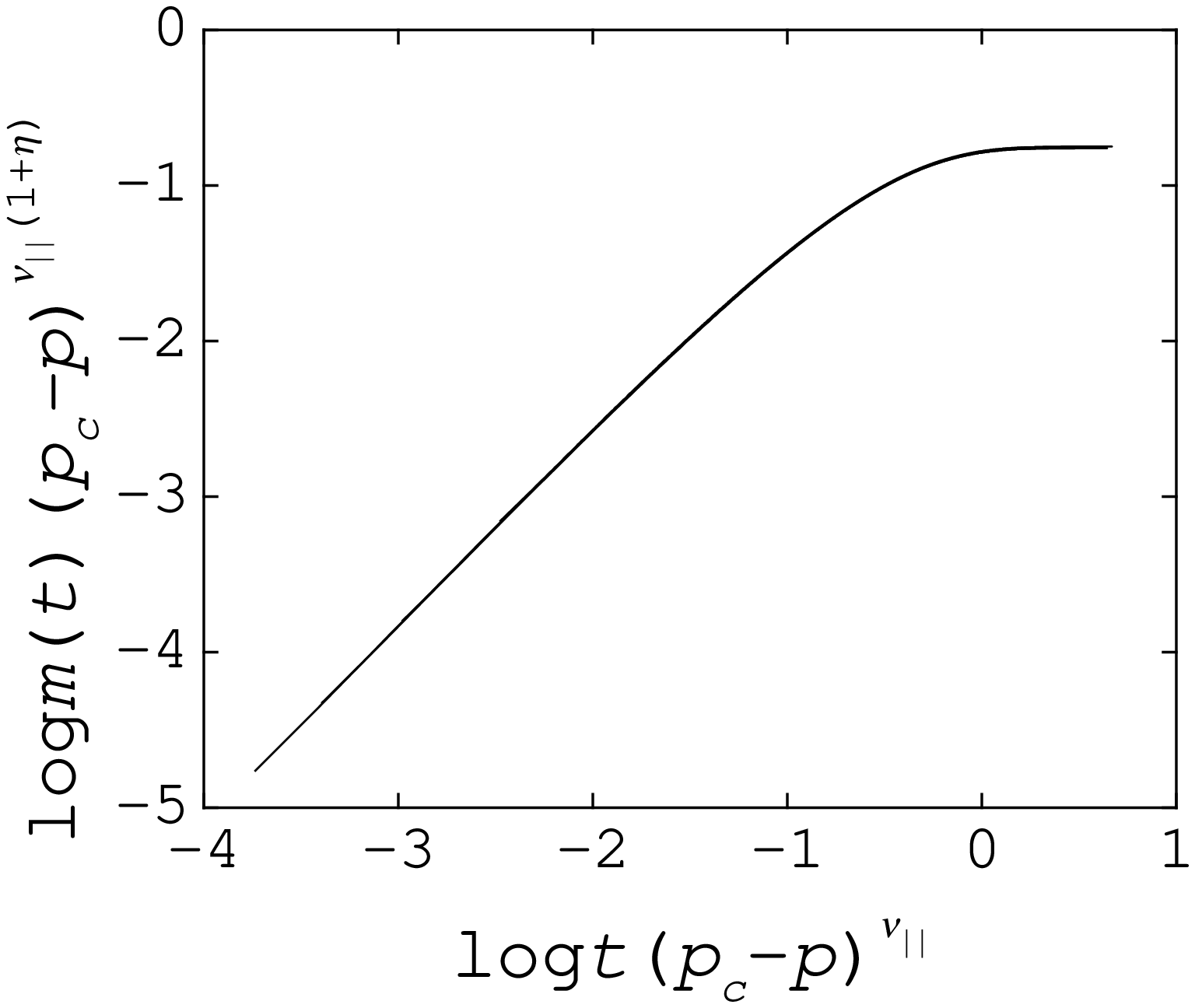,width=3.5in,height=3in}}
\caption{Data collapse plot for $m(t)$ in the subcritical
state for $p=0.670$, 0.688, 0.696 and 0.700.\\}
\label{fig:4}
\end{figure}

\vskip 0.5in

\begin{figure}\narrowtext
\centerline{\psfig{figure=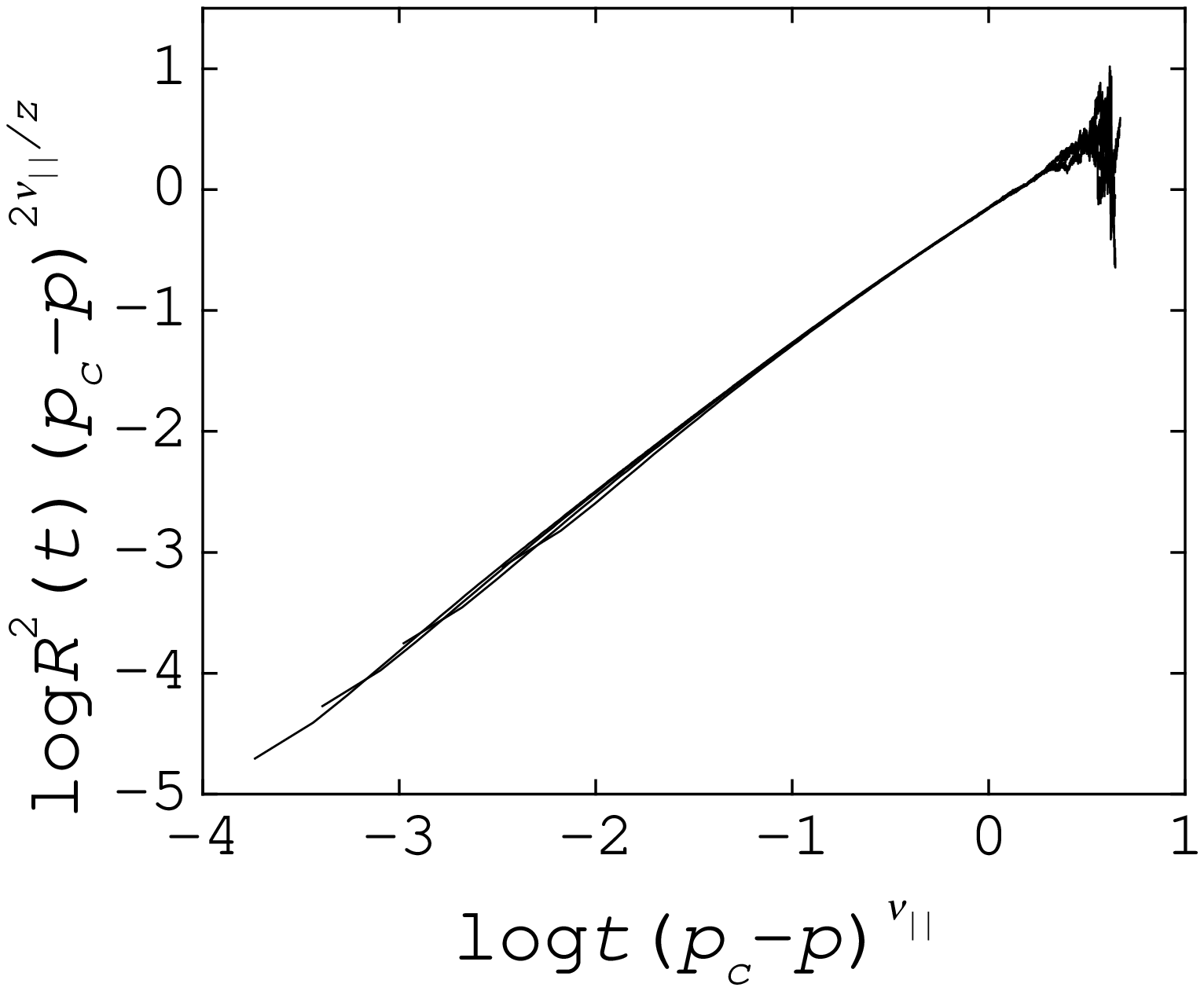,width=3.5in,height=3in}}
\caption{Data collapse plot for $R^2(t)$ in the subcritical
state for $p=0.670$, 0.688, 0.696 and 0.700.}
\label{fig:5}
\end{figure}

\newpage

\vskip 0.5in

\vskip 0.5in

\end{multicols}

\end{document}